\documentclass[lettersize,journal]{IEEEtran}

\usepackage{amsmath,amsfonts}
\usepackage{algorithmic}
\usepackage{algorithm}
\usepackage{array}
\usepackage{textcomp}
\usepackage{url}
\usepackage{verbatim}
\usepackage{graphicx}
\usepackage{cite}
\usepackage{subcaption}
\usepackage{xcolor}

\usepackage{mathtools} 

\ifodd 0
\newcommand{\EditMush}[1]{\textcolor[rgb]{1.0,0.0,0.0}{#1}}
\else
\newcommand{\EditMush}[1]{#1}
\fi

\usepackage{tikz}
\newcommand\copyrighttext{%
  \footnotesize \copyright~2026 IEEE. Personal use permitted. Submitted to IEEE Wireless Communications Letters. Copyright may be transferred without notice.}
\newcommand\copyrightnotice{%
\begin{tikzpicture}[remember picture,overlay]
\node[anchor=south, yshift=15pt] at (current page.south) {\copyrighttext};
\end{tikzpicture}%
}

\begin{document}


\title{Platform-Aware Channel Knowledge Mapping via Mutual Antenna Pattern Learning in 3D Wireless Links}

\author{Mushfiqur Rahman, \IEEEmembership{Graduate Student Member, IEEE}, \.{I}smail G\"{u}ven\c{c}, \IEEEmembership{Fellow, IEEE}, Jason A. Abrahamson, and Arupjyoti Bhuyan, \IEEEmembership{Senior Member, IEEE}
\thanks{We use measurement data~\cite{masrur2025collection,IEEEDataPort_4} published by the NSF Aerial Experimentation and Research Platform on Advanced Wireless (AERPAW)~\cite{9061157} platform. This research is supported in part by the NSF award CNS-2332835 and CNS-2450593, and the INL Laboratory Directed Research Development (LDRD) Program under BMC No. 264247, Release No. 26 on BEA’s Prime Contract No. DE-AC07-05ID14517.
\textit{(Corresponding author: {I}smail~G\"{u}ven\c{c}.)}}
\thanks{Mushfiqur Rahman and \.{I}smail G\"{u}ven\c{c} are with the Department of Electrical and Computer Engineering, North Carolina State University, Raleigh, NC 27606, USA (e-mail: mrahman7@ncsu.edu; iguvenc@ncsu.edu).}
\thanks{Jason A. Abrahamson and Arupjyoti Bhuyan are with the Idaho National Laboratory, Idaho Falls, ID 83402, USA (e-mail: jason.abrahamson@inl.gov; arupjyoti.bhuyan@inl.gov).}
}

\markboth{
}
{Rahman \MakeLowercase{\textit{et al.}}: Characterization of the Combined Effective Radiation Pattern of UAV-Mounted Antennas and Ground Station}


\maketitle
\copyrightnotice

\begin{abstract}
\EditMush{This letter proposes a platform-aware framework to characterize wireless links by empirically modeling the `near-platform' scattering and reflections induced by the hardware mounting structures of both endpoints. We model the link characteristics as a novel mutual antenna pattern—a joint function of the angle of arrival (AoA) and angle of departure (AoD). We demonstrate that while individual platform-aware patterns are mathematically unidentifiable from power measurements, the coupled mutual pattern can be effectively estimated in a least-squares sense. Our framework is evaluated using noisy measurement data, revealing that as few as 10 measurements per joint-angular bin are sufficient. The proposed methodology is validated through cross-validation of experimental subsets, demonstrating that the learned mutual radiation pattern reduces path loss estimation errors by up to 10~dB compared to traditional models using isolated anechoic chamber antenna gains.}
\end{abstract}


\section{Introduction}
\EditMush{Sixth-generation (6G) wireless networks are defined by a critical transition from environment-unaware systems to fully environment-aware communication paradigms. Unlike fifth-generation (5G) frameworks that rely on costly real-time beam sweeping and channel state information (CSI) feedback~\cite{wu2025ckmimagenet}, 6G networks aim to exploit an abundance of location-tagged data to perceive the local wireless environment. Central to this vision is the \textit{channel knowledge map} (CKM), a site-specific database that provides intrinsic channel properties such as region-specific path loss parameters, propagation delays, and multi-path angles of arrival/departure (AoA/AoD)~\cite{li2022channel}. However, while traditional CKMs focus on environment-centric data (e.g., building layouts and satellite maps), we argue that \textit{platform-specific knowledge} represents a vital, yet overlooked, dimension. The physical structure of integrated platforms, such as the airframe of an uncrewed aerial vehicle (UAV) or the chassis of a ground vehicle, intrinsically shapes the wireless link, necessitating a move toward platform-aware channel characterization exploiting their physical properties.}

\EditMush{We propose that the interaction between the physical structures of two nodes gives rise to a \textit{mutual antenna pattern}, where the 3D link characteristics are defined as a joint radiation pattern function of both the AoA and AoD simultaneously. This approach addresses a key limitation of existing location-specific CKMs: the inability to account for the 3D orientation of mobile terminals, which significantly alters structurally-induced scattering and reflection. In 3D wireless links involving UAVs or high-altitude platforms (HAPs), which are vital to 6G networks, 3D rotations can create dynamic multipath profiles~\cite{wu2023environment}. Additionally, the proposed approach can alleviate the need for frequent beam scanning in massive multiple-input multiple-output (MIMO) systems by predicting precise beamforming gains using the mutual antenna pattern.}
\begin{figure}[!t]
\centerline{\includegraphics[width=\linewidth,trim={6.8cm 4.2cm 6.5cm 4.8cm},clip]{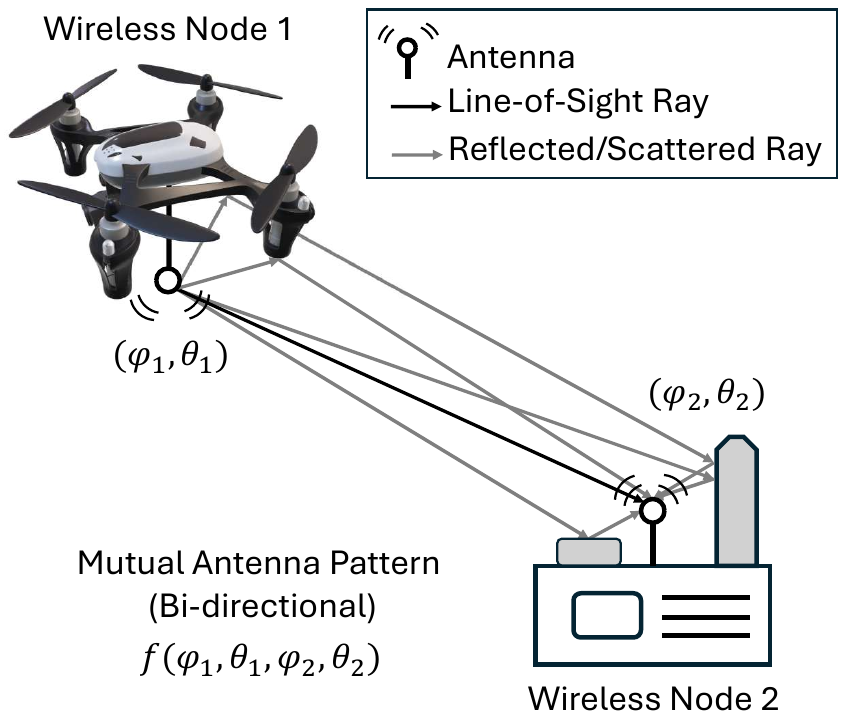}}
\caption{\EditMush{Wireless link between two nodes with integrated physical structures. Reflections and scattering from the node bodies create multipath components that depend on their 3D orientations and the departure/arrival angles at both terminals.}}
\label{fig:link_def}
\vspace{-4mm}
\end{figure}

\section{Proposed Framework of Channel Modeling}
\subsection{\EditMush{Structurally-Induced Wireless Link Environments}}
\label{sec:env_def}
\EditMush{A wireless link is fundamentally defined as the electromagnetic communication path between a transmitter and a receiver. In practical deployments, these terminals are rarely isolated; instead, they are integrated into physical platforms such as UAVs, uncrewed ground vehicles (UGVs), or satellites. The physical structure of these nodes intrinsically shapes the wireless link through various mechanisms, including signal blockage (e.g., airframe shadowing), reflections, and scattering from the node's body, as illustrated in Fig.~\ref{fig:link_def}.}

\EditMush{In the absence of external obstructions, the wireless channel characteristic is primarily governed by the point-to-point vector connecting the two nodes, commonly referred to as the line-of-sight (LoS) direction. This LoS vector, when mapped to the local reference frames of the respective nodes, defines the AoD and AoA. We denote the AoD at Node 1 as $\omega_1 = (\phi_1, \theta_1)$ and the AoA at Node 2 as $\omega_2 =(\phi_2, \theta_2)$, where $\phi$ and $\theta$ represent the azimuth and elevation angles, respectively. Crucially, these angular coordinates are a joint function of the absolute LoS vector and the 3D orientation (roll, pitch, and yaw) of the platforms~\cite{craig2018introduction}. Next, we define the concept of mutual antenna pattern based on this geometric foundation.}

\subsection{\EditMush{Mutual Antenna Pattern}}
\label{sec:mutual_pat}
\EditMush{Let the complex radiation patterns of Node 1 and Node 2, in the presence of their respective physical platforms (e.g., as characterized in an anechoic chamber), be denoted by $G_1(\omega_1)$ and $G_2(\omega_2)$. Here, $\omega_1$ and $\omega_2$ represent the respective AoD and AoA as specified in Section~\ref{sec:env_def}. We introduce the \textit{mutual antenna pattern} as the joint gain function resulting from the interaction of both platforms. In a linear scale, this is expressed as the product of the individual directional gains as follows:}
\EditMush{
\begin{equation}
    G_\text{m}(\omega_1, \omega_2) = G_1(\omega_1) G_2(\omega_2) \equiv G_\text{m}(\Omega),
\label{eq:basic_mutual_def}
\end{equation}}
\EditMush{where $\Omega = (\omega_1, \omega_2)$ denotes the joint angular state comprising both AoD and AoA. Notably, the multiplicative relationship in~\eqref{eq:basic_mutual_def} corresponds to an additive operation in the logarithmic domain, facilitating a linear formulation for estimation.}

\EditMush{To establish the physical significance of \eqref{eq:basic_mutual_def}, we consider its role in simplifying the characterization of multipath propagation. Generally, a wireless channel can be modeled as the superposition of $M$ multipath components:}
\EditMush{\begin{equation}
    \vec{r}(\omega_1, \omega_2, d_{\mathrm{3D}}) = \sum_{i=1}^{M} a_i \exp(j \psi_i),
\label{eq:total_signal_effective}
\end{equation}}
\hspace{-1mm}\EditMush{where $\vec{r}$ is the received signal phasor, $d_{\mathrm{3D}}$ is the 3D Euclidean distance between nodes, and $a_i$ and $\psi_i$ represent the magnitude and phase of the $i$-th multipath ray, respectively.}

\EditMush{The mutual antenna pattern effectively encapsulates the aggregate effect of all platform-induced reflections and scattering into a unified gain term. Consequently, the received power can be modeled using a modified free-space path loss (FSPL) formulation, which consolidates the individual antenna gains into a single joint term as follows:}
\EditMush{\begin{equation}
    P_r(\omega_1, \omega_2, d_{\mathrm{3D}}) = \left( \frac{\lambda}{4 \pi d_{\mathrm{3D}}} \right)^2 G_\text{m}(\omega_1, \omega_2),
\label{eq:alternate_def}
\end{equation}}
\EditMush{where $P_r = |\vec{r}|^2$ is the received power in Watts and $\lambda$ is the operating wavelength. Thus, from \eqref{eq:alternate_def}, the mutual antenna gain is defined as the normalized power involving the vector sum of all multi-ray components interacting with the two nodes' physical structures.} 

\subsection{\EditMush{Empirical Estimation of \(G_\text{m}\) from Noisy RSS Measurements}}
\label{sec:pro_method}
\EditMush{Researchers have traditionally characterized platform-induced radiation patterns using anechoic chamber measurements~\cite{semkin2021lightweight, catherwood2019radio, badi2020experimentally} with spectrum or vector network analyzers~\cite{balderas2019low}. Additionally, electromagnetic simulations using tools like Altair FEKO~\cite{rizwan2017impact} and CST Microwave Studio~\cite{balderas2019low} have been widely employed.}
\EditMush{Despite these advancements, practical wireless deployments often necessitate in-situ characterization following platform integration. Unlike laboratory settings, however, field measurement data is typically less ideal. It contains node positions and received signal strength (RSS) measurements subject to hardware noise, motor vibrations, and external environmental factors, including shadowing and small-scale fading. Furthermore, these observations often suffer from sparse angular coverage. In this work, we estimate $G_\text{m}(\omega_1,\omega_2)$ using a least-squares (LS) approach as follows.} 

\EditMush{Let the 3D angular space $\omega$ be discretized into $K$ unique directions: $\{\omega^{(1)}, \dots, \omega^{(K)}\}$. If we attempt to estimate the individual patterns $G_1(\omega)$ and $G_2(\omega)$ as independent variables, we must solve for $2K$ unknowns. For $n$ observations, the system of linear equations in the dB scale is:}
\EditMush{\begin{equation}
\small
\label{eqn:estimation_separate}
\mathbf{C} \mathbf{g} = \mathbf{s} \implies
\begin{bmatrix}
C_{1,1} & \cdots & C_{1,2K} \\
C_{2,1} & \cdots & C_{2,2K} \\
\vdots & \ddots & \vdots \\
C_{n,1} & \cdots & C_{n,2K}
\end{bmatrix}
\begin{bmatrix}
G_1(\omega^{(1)}) \\ \vdots \\ G_1(\omega^{(K)}) \\ G_2(\omega^{(1)}) \\ \vdots \\ G_2(\omega^{(K)})
\end{bmatrix} = 
\begin{bmatrix}
S_1 \\ S_2 \\ \vdots \\ S_n
\end{bmatrix},
\end{equation}}
\EditMush{where $\mathbf{s}$ is the vector of noisy observations representing normalized received power in dB (with the normalization term given in \eqref{eq:alternate_def}), $\mathbf{g}$ is the vector of unknown gains to be estimated, and $\mathbf{C}$ is the coefficient matrix. The structure of $\mathbf{C}$ satisfies the constraints $\sum_{j=1}^{K} C_{i,j} = 1$ and $\sum_{j=K+1}^{2K} C_{i,j} = 1$ for every observation $i$. Physically, these constraints imply that for a given link direction $\Omega = (\omega_1, \omega_2)$, exactly one directional bin is active for each respective antenna. However, due to these constraints, the matrix $\mathbf{C}$ is rank-deficient. Specifically, the sum of the first $K$ columns minus the sum of the remaining $K$ columns yields a zero vector. This indicates that $G_1(\omega)$ and $G_2(\omega)$ are not individually identifiable from power measurements alone, as their gains are intrinsically merged.}

\EditMush{To resolve this, we discretize the mutual antenna pattern $G_\text{m}(\Omega)$ into $K^2$ joint angular bins: $\Omega^{(1)}, \dots, \Omega^{(K^2)}$. The estimation problem for $K^2$ variables is formulated as follows:}
\EditMush{\begin{equation}
\small
\label{eqn:estimation_joint}
\begin{bmatrix}
A_{1,1} & \cdots & A_{1,K^2} \\
\vdots & \ddots & \vdots \\
A_{n,1} & \cdots & A_{n,K^2}
\end{bmatrix}
\begin{bmatrix}
G_\text{m}(\Omega^{(1)}) \\ \vdots \\ G_\text{m}(\Omega^{(K^2)})
\end{bmatrix} = 
\begin{bmatrix}
S_1 \\ \vdots \\ S_n
\end{bmatrix},
\end{equation}}
\EditMush{where $\sum_{j=1}^{K^2} A_{i,j} = 1$. In this formulation, the matrix is no longer inherently rank-deficient. Since each row contains exactly one active coefficient, the global LS problem decomposes into $K^2$ independent single-variable problems. The solution for each joint direction $\Omega^{(k)}$ simplifies to the sample mean of the normalized observations falling within that angular bin:}
\EditMush{\begin{equation}
\hat{G}_\text{m}(\Omega^{(k)}) = \frac{1}{N_k} \sum_{i=1}^{N_k} S_i{(\Omega^{(k)})},
\label{eqn:combined_learning}
\end{equation}}
\EditMush{where $N_k$ denotes the number of observations in the subset $S{(\Omega^{(k)})}\subset \mathbf{s}$ associated with the $k$-th joint angular bin $\Omega^{(k)}$.}

\section{Experimental Results}
\subsection{Datasets and Experimental Configurations}
\label{sec:datasets}
We employ two publicly available wireless link datasets provided by the NSF Aerial Experimentation and Research Platform on Advanced Wireless (AERPAW)~\cite{9061157}. The characteristics of these datasets are summarized in Table~\ref{tab:exp_list}. Dataset 1 consists of wireless links between a UAV and a UGV node, while Dataset 2 employs a link between a base station tower and a UAV. In Table~\ref{tab:exp_list}, experiments A1-A5 use a UGV as a transmitter with an antenna height of 1.5 meters, while experiments B1-B5 use a tower with an antenna height of 10 meters. \EditMush{Both experiments were conducted in a rural environment with an unobstructed LoS, except for structural shadowing and blockages caused by the nodes' platforms.}

\EditMush{For a robust evaluation, we select specific experiments for training to estimate the mutual antenna pattern $\hat{G}_\text{m}(\Omega)$. We discretize the angular space $\omega$ using $90^\circ$ azimuth bins and $5^\circ$ elevation bins, totaling $36 \times 4 = 144$ angular bins for each node, and consequently $144^2 = 20,736$ joint angular bins for $\Omega$. To mimic a practical scenario with limited data availability, we cap the number of observations per joint angular bin at $N_k = 10$. These $N_k$ samples are selected randomly from the available training data for each $\Omega$. Due to the high dimensionality of the joint angular space, many bins contain fewer than $N_k$ observations or are entirely empty. Such bins are marked as unknown in $\hat{G}_\text{m}(\Omega)$.}
\renewcommand{\arraystretch}{1.3} 
\begin{table}[ht]
\centering
\caption{Experiments considered in this study.}
\begin{tabular}{ccccc}
\hline
\textbf{Dataset} & \textbf{Label} & \textbf{UAV Altitude} & \textbf{Bandwidth}  & \textbf{Carrier}                                \\ \hline
Dataset 1 & A1                         & 37~m               & 125~kHz              & 3.32~GHz \\ \hline
Dataset 1 & A2                         & 47~m               & 125~kHz              & 3.32~GHz \\ \hline   
Dataset 1 & A3                         & 21~m               & 125~kHz              & 3.32~GHz \\ \hline 
Dataset 1 & A4                         & 28~m               & 125~kHz            & 3.32~GHz \\ \hline 
Dataset 1 & A5                         & 28~m               & 125~kHz               & 3.32~GHz \\ \hline \hline
Dataset 2 & B1                         & 40~m               & 5~MHz              & 3.3~GHz \\ \hline
Dataset 2 & B2                         & 70~m               & 5~MHz              & 3.3~GHz \\ \hline   
Dataset 2 & B3                         & 100~m               & 5~MHz              & 3.3~GHz  \\ \hline
Dataset 2 & B4                         & 40~m               & 1.25~MHz              & 3.3~GHz \\ \hline
Dataset 2 & B5                         & 40~m               & 2.5~MHz              & 3.3~GHz \\ \hline
\end{tabular}
\label{tab:exp_list}
\end{table}

\subsection{\EditMush{Baselines and Evaluation Metrics}}
\EditMush{To evaluate the proposed framework, we compare the learned mutual antenna pattern against the anechoic chamber measurements. These are the radiation patterns of the individual antennas without the structural effects of their platforms. This comparison contrasts in-situ characterization with lab-based measurements, especially in scenarios where anechoic chamber testing is impractical due to platform size (e.g., UGVs), or the need for frequent reconfiguration.}

\EditMush{When the isolated antenna gains are used in~\eqref{eq:alternate_def}, the model effectively reduces to the standard FSPL formulation. Performance is quantified using the mean absolute error (MAE) of the RSS estimation on a held-out test experiment. Notably, since the mutual antenna gain is platform-specific, the physical configuration of the nodes must remain identical between the training and testing phases. Consequently, cross-dataset evaluation (e.g., testing on Dataset 2 using a model trained on Dataset 1) is not feasible, as these datasets involve using different antennas and structural platforms.}

\subsection{\EditMush{Snapshot of Reconstructed Mutual Antenna Gain}}
\begin{figure}[t!]
    \centering
    \begin{subfigure}{0.23\textwidth}
        \centering
        \includegraphics[width=\linewidth,trim={6.1cm 9.6cm 6.1cm 9.0cm},clip]{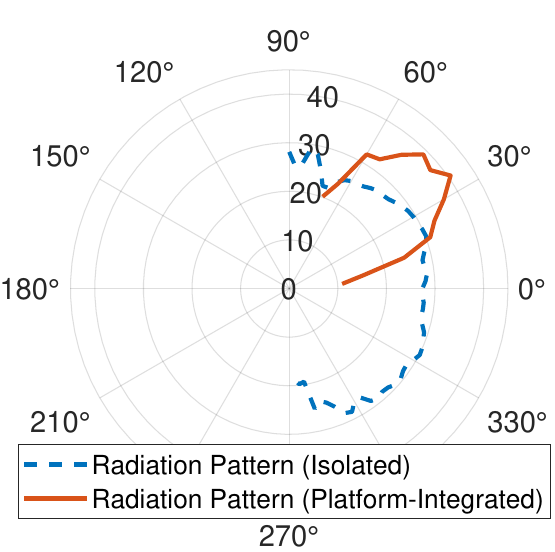}
        \caption{A2.}
        \label{fig:sub1}
    \end{subfigure}
    \begin{subfigure}{0.23\textwidth}
        \centering
        \includegraphics[width=\linewidth,trim={6.1cm 9.6cm 6.1cm 9.0cm},clip]{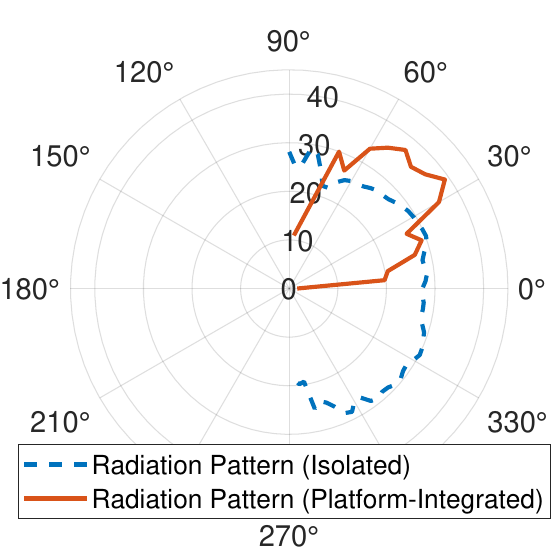}
        \caption{A4, A5.}
        \label{fig:sub1}
    \end{subfigure}
    \begin{subfigure}{0.23\textwidth}
        \centering
        \includegraphics[width=\linewidth,trim={6.1cm 9.6cm 6.2cm 9.0cm},clip]{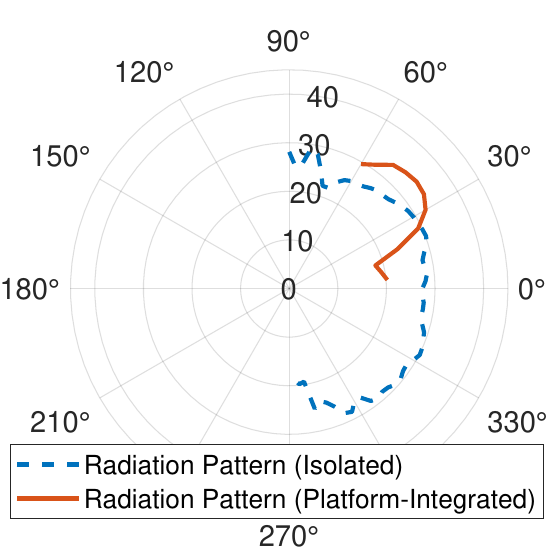}
        \caption{B2.}
        \label{fig:sub1}
    \end{subfigure}
    \begin{subfigure}{0.23\textwidth}
        \centering
        \includegraphics[width=\linewidth,trim={6.1cm 9.6cm 6.2cm 9.0cm},clip]{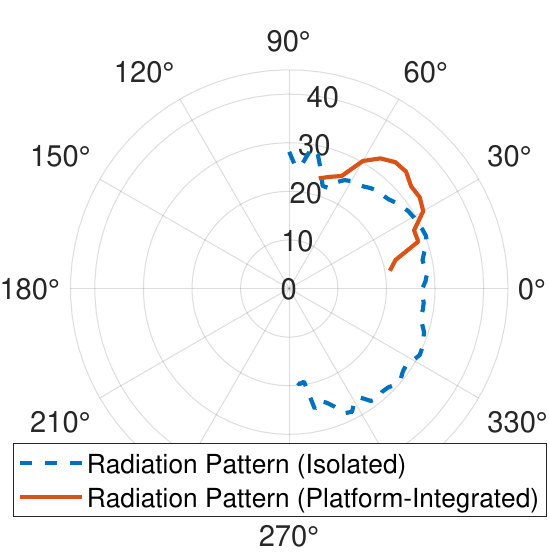}
        \caption{B3.}
        \label{fig:sub1}
    \end{subfigure}
    \vspace{-1mm}
    \caption{\EditMush{Snapshot of reconstructed mutual antenna gain as a function of elevation angle. Subfigures (a)--(d) denote the experimental training subsets in Table~\ref{tab:exp_list}.}}
    \label{fig:ant_patt_elev}\vspace{-4mm}
\end{figure}

\EditMush{The reconstructed mutual antenna pattern $G_\text{m}(\Omega)$ is a four-dimensional function, as both the AoD and AoA comprise an azimuth and an elevation angle. To visualize these joint gains, we evaluate $G_\text{m}(\theta_1, \phi_1, \theta_2, \phi_2)$ and average it over the azimuth $\phi_1$ to obtain the elevation radiation pattern for $\theta_1$, where $\theta_2 = \theta_1 + c_1$, $\phi_2 = \phi_1 + c_2$, and $c_1, c_2$ are fixed angular offsets. These patterns, shown in Fig.~\ref{fig:ant_patt_elev}, do not represent the transmitter or receiver antennas individually, but rather the integrated gain of both nodes in specific coupled directions.}
\EditMush{For comparison, the dashed curves represent the combined anechoic chamber patterns of the two isolated antennas. As observed in Fig.~\ref{fig:ant_patt_elev}(a) and Fig.~\ref{fig:ant_patt_elev}(b) for Dataset 1 and Fig.~\ref{fig:ant_patt_elev}(c) and Fig.~\ref{fig:ant_patt_elev}(d) for Dataset 2, the physical platforms induce a unique radiation profile that diverges significantly from the baseline summation of isolated antenna gains.}

\subsection{RSS Estimation Performance}
\label{sec:comparison}
In this subsection, we evaluate the performance of RSS estimation using the proposed mutual antenna gain \(G_\text{m}(\Omega)\).
\begin{figure*}[!t]
    \centering

    \begin{subfigure}{0.32\textwidth}
        \centering
        \includegraphics[width=\linewidth,trim={4.3cm 10.25cm 5.1cm 10.6cm},clip]{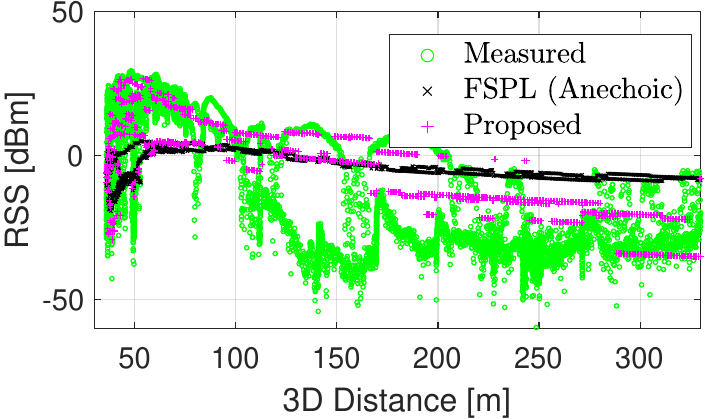}
        \caption{A1, Trained on A2.}
        \label{fig:sub1}
    \end{subfigure}
    \begin{subfigure}{0.32\textwidth}
        \centering
        \includegraphics[width=\linewidth,trim={4.3cm 10.25cm 5.1cm 10.6cm},clip]{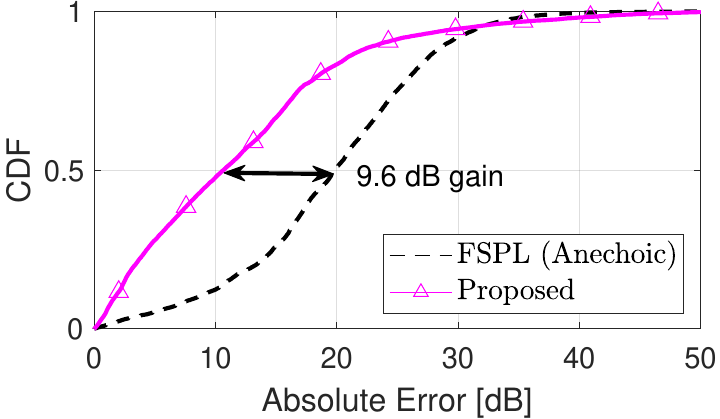}
        \caption{A1, Trained on A2.}
        \label{fig:sub1}
    \end{subfigure}
    \begin{subfigure}{0.32\textwidth}
        \centering
        \includegraphics[width=\linewidth,trim={4.4cm 10.0cm 4.1cm 10.35cm},clip]{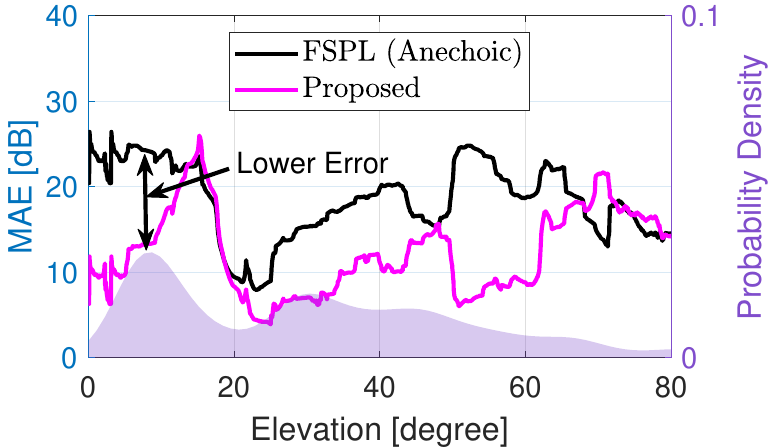}
        \caption{A1, Trained on A2.}
        \label{fig:sub1}
    \end{subfigure}
    \begin{subfigure}{0.32\textwidth}
        \centering
        \includegraphics[width=\linewidth,trim={4.3cm 10.25cm 5.1cm 10.6cm},clip]{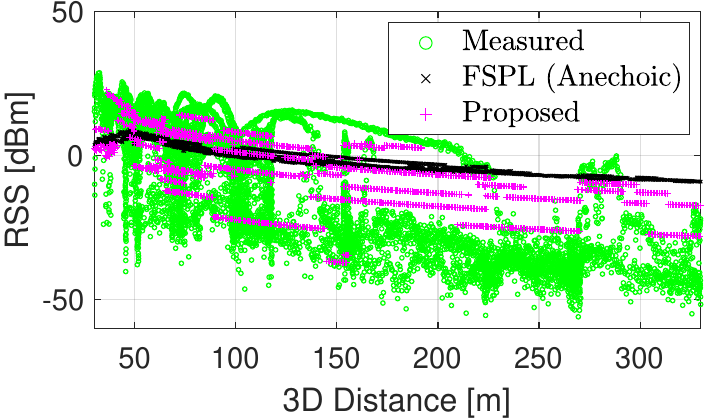}
        \caption{A3, Trained on A4, A5.}
        \label{fig:sub1}
    \end{subfigure}
    \begin{subfigure}{0.32\textwidth}
        \centering
        \includegraphics[width=\linewidth,trim={4.3cm 10.25cm 5.1cm 10.6cm},clip]{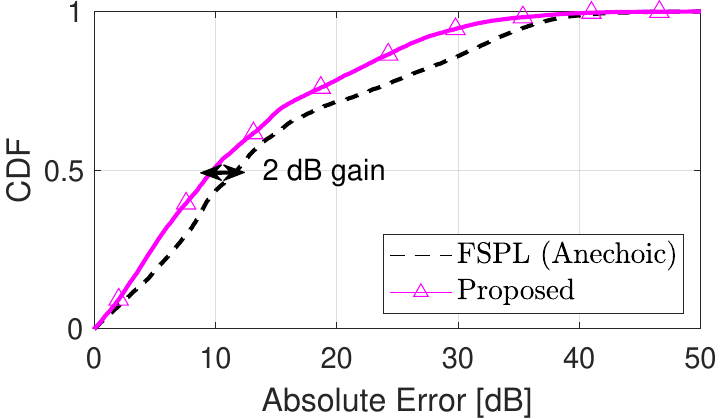}
        \caption{A3, Trained on A4, A5.}
        \label{fig:sub1}
    \end{subfigure}
    \begin{subfigure}{0.32\textwidth}
        \centering
        \includegraphics[width=\linewidth,trim={4.4cm 10.0cm 4.1cm 10.35cm},clip]{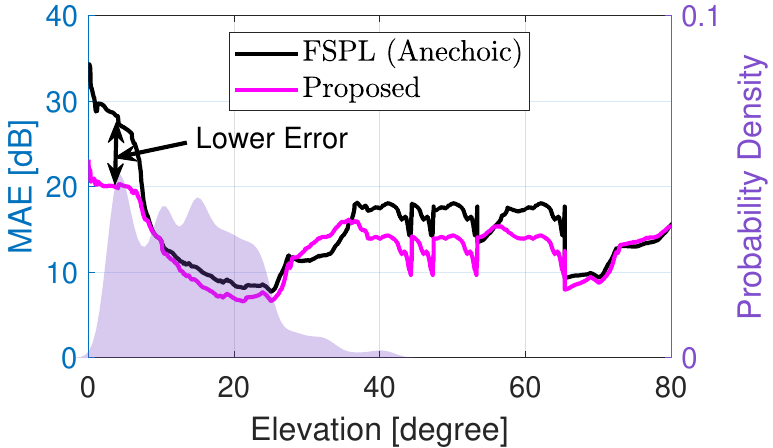}
        \caption{A3, Trained on A4, A5.}
        \label{fig:sub1}
    \end{subfigure}
    \vspace{-1mm}
        \caption{RSS prediction performance for \EditMush{test experiments} A1 and A3: \EditMush{comparison between baseline and proposed method}. (a) and (d) \EditMush{spatial tracking of measured and predicted RSS across 3D distance}; (b) and (e) CDF of absolute error; (c) and (f) \EditMush{MAE as a function of elevation angle alongside the elevation angle probability distribution.}}
    \label{fig:comparison_dataset1}\vspace{-4mm}
\end{figure*}
Fig.~\ref{fig:comparison_dataset1} illustrates the performance comparison between the anechoic chamber gain and the estimated \(\hat{G}_\text{m}(\Omega)\) for the two train-test scenarios of Dataset 1. Fig.~\ref{fig:comparison_dataset1}(a) and Fig.~\ref{fig:comparison_dataset1}(d) show the measured and predicted RSS from both methods across 3D distance. The derived gain captures more variations in the measurements. Specifically, at larger 3D distances, which correspond to lower elevation angles, the predicted RSS using the proposed method falls below the one from anechoic chamber gain, \EditMush{capturing structural shadowing}, and matches better with the actual measurements. Additionally, at smaller 3D distances, the proposed method's predicted power is higher, aligning more closely with the measurements. However, the model does not capture periodic variations with distance, since the antenna pattern does not account for distance.

Fig.~\ref{fig:comparison_dataset1}(c) and Fig.~\ref{fig:comparison_dataset1}(f) display the comparative performance of the two methods across elevation angles. The shaded regions in these plots represent the test data's probability density of elevation angles for the test experiment. The proposed method outperforms the baseline for both lower and higher elevation angles up to~\(80^\circ\). Specifically, the performance increases by more than \(10\)~dB for elevation angles ranging from \(0^\circ\) to \(10^\circ\), shows near-zero gain around \(20^\circ\), and then exhibits a subsequent \(\sim 10\)~dB improvement near \(40^\circ\). This pattern is consistent with the elevation radiation patterns presented in Fig.~\ref{fig:ant_patt_elev}(a) and Fig.~\ref{fig:ant_patt_elev}(b), which demonstrate nearly identical antenna gains for both the \(G_\text{m}(\Omega)\) and anechoic chamber-based gain around the \(20^\circ\) elevation angle. The cumulative density function (CDF) plots of absolute error for both methods are shown in Fig.~\ref{fig:comparison_dataset1}(b) and Fig.~\ref{fig:comparison_dataset1}(e), demonstrating the median performance improvement of \(9.6\)~dB and \(2.0\)~dB, respectively. 

\begin{figure*}[!t]
    \centering
    \begin{subfigure}{0.32\textwidth}
        \centering
        \includegraphics[width=\linewidth,trim={4.3cm 10.25cm 5.1cm 10.6cm},clip]{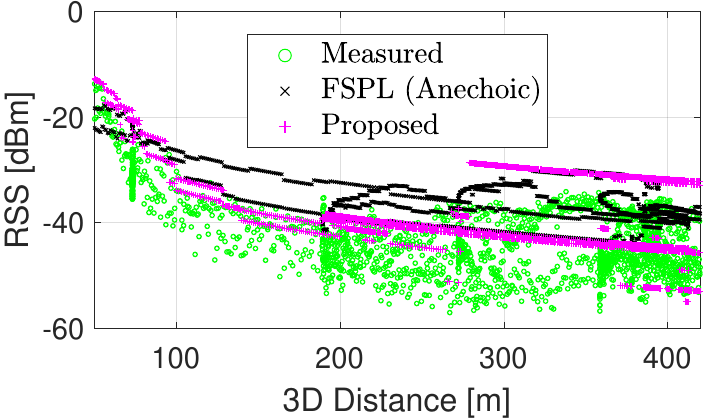}
        \caption{B1, Trained on B2.}
        \label{fig:sub1}
    \end{subfigure}
    \begin{subfigure}{0.32\textwidth}
        \centering
        \includegraphics[width=\linewidth,trim={4.3cm 10.25cm 5.1cm 10.6cm},clip]{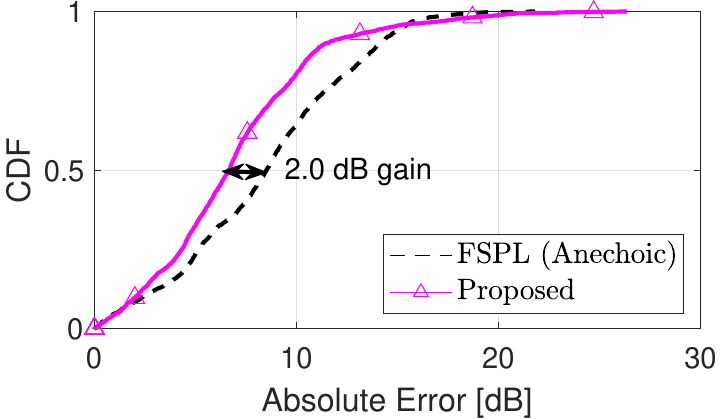}
        \caption{B1, Trained on B2.}
        \label{fig:sub1}
    \end{subfigure}
    \begin{subfigure}{0.32\textwidth}
        \centering
        \includegraphics[width=\linewidth,trim={4.4cm 10.0cm 4.1cm 10.35cm},clip]{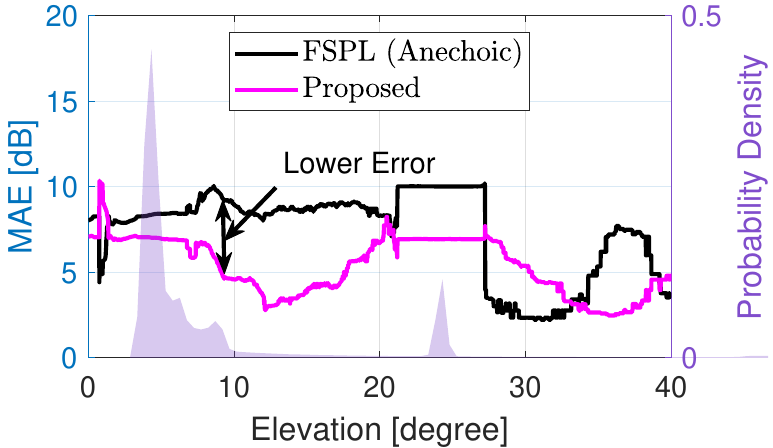}
        \caption{B1, Trained on B2.}
        \label{fig:sub1}
    \end{subfigure}
    \begin{subfigure}{0.32\textwidth}
        \centering
        \includegraphics[width=\linewidth,trim={4.3cm 10.25cm 5.1cm 10.6cm},clip]{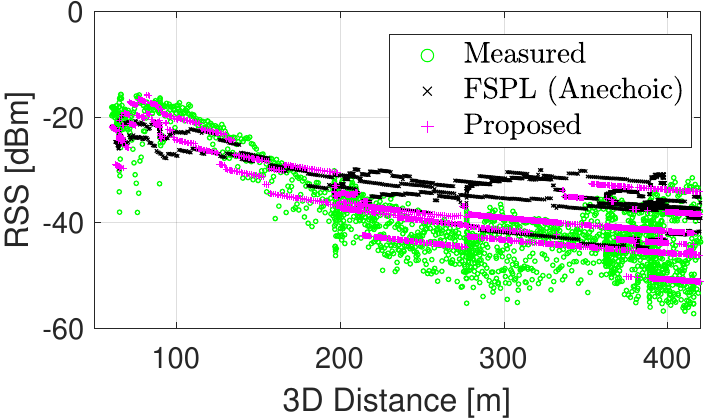}
        \caption{B2, Trained on B3.}
        \label{fig:sub1}
    \end{subfigure}
    \begin{subfigure}{0.32\textwidth}
        \centering
        \includegraphics[width=\linewidth,trim={4.3cm 10.25cm 5.1cm 10.6cm},clip]{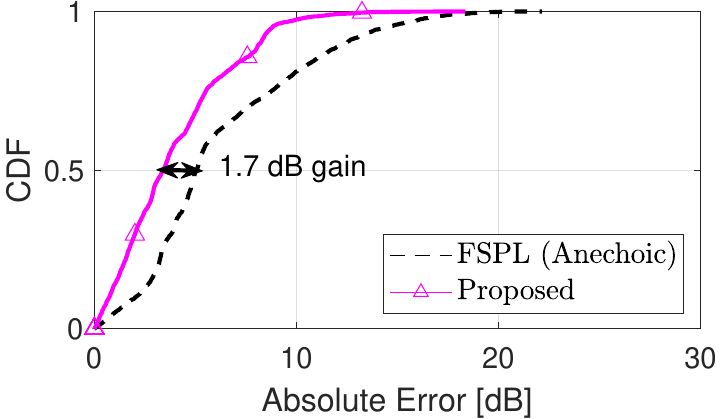}
        \caption{B2, Trained on B3.}
        \label{fig:sub1}
    \end{subfigure}
    \begin{subfigure}{0.32\textwidth}
        \centering
        \includegraphics[width=\linewidth,trim={4.4cm 10.0cm 4.1cm 10.35cm},clip]{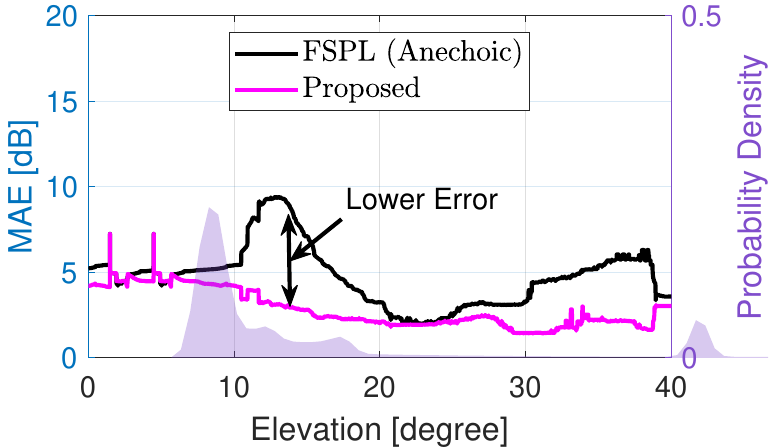}
        \caption{B2, Trained on B3.}
        \label{fig:sub1}
    \end{subfigure}
    \vspace{-1mm}
        \caption{RSS prediction performance for \EditMush{test experiments} B1 and B2: \EditMush{comparison between baseline and proposed method}. (a) and (d) \EditMush{spatial tracking of measured and predicted RSS across 3D distance}; (b) and (e) CDF of absolute error; (c) and (f) \EditMush{MAE as a function of elevation angle alongside the elevation angle probability distribution.}}
    \label{fig:comparison_dataset2}\vspace{-4mm}
\end{figure*}

\renewcommand{\arraystretch}{1.3} 
\begin{table}[t!]
\centering
\caption{RSS modeling error.}
\begin{tabular}{ccccc}
\hline
\textbf{Test} & \textbf{Train} & \textbf{MAE (Anechoic)} & \textbf{MAE (Proposed)}  & \textbf{Gain}                              \\ \hline
A1 & A2                         & 19.47~dB               & 12.09~dB  & 7.38~dB             \\ \hline

A3 & A4, A5                        & 14.85~dB               & 12.24~dB &  2.61~dB           \\ \hline   

B1 & B2                         & 8.41~dB               & 6.99~dB &  1.42~dB            \\ \hline 

B2 & B3                         & 6.28~dB              & 3.95~dB   & 2.33~dB         \\ \hline

B4 & B3                         & 7.51~dB              & 4.91~dB  &  2.60~dB          \\ \hline
B5 & B3                         & 7.43~dB              & 4.42~dB &  3.01~dB            \\ \hline
B1 & B3                         & 8.41~dB              & 5.14~dB &  3.27~dB        \\ \hline
\end{tabular}
\label{tab:per_compr}
\end{table}
\renewcommand{\arraystretch}{1.3} 
\begin{table}[!]
\centering
\caption{Performance gain vs. bandwidth.}
\begin{tabular}{ccccc}
\hline
\textbf{Bandwidth} [MHz] & \textbf{0.125} & \textbf{1.25} & \textbf{2.5}  & \textbf{5}                              \\ \hline
MAE (Anechoic) [dB] & 19.47 & 7.51 & 7.43  & 6.28       \\ \hline
MAE (Proposed) [dB] & 12.09 & 4.91 & 4.42  & 3.95       \\ \hline
\textbf{Performance Gain} [dB] & \textbf{7.38} & \textbf{2.60} & \textbf{3.01}  & \textbf{2.33}       \\ \hline
\end{tabular}
\label{tab:per_compr2}
\vspace{-4mm}
\end{table}
Similarly, a performance gain is observed for Dataset 2, as shown in Fig.~\ref{fig:comparison_dataset2}. The median improvement for the two train-test scenarios is $2.0$~dB and $1.7$~dB, respectively, which is notably lower than the gains observed in Dataset 1. \EditMush{This disparity suggests that in Dataset 1, the UGV's physical structure and nearby objects significantly influenced the link. In contrast, one node in Dataset 2 was a base station with an antenna mounted at a higher altitude and fewer surrounding structural components, thereby limiting the potential gains from platform-aware pattern learning.}
Table~\ref{tab:per_compr} presents the performance gains obtained from the conducted experiments.

\section{Discussion}
\EditMush{The proposed method demonstrates robustness to the number of observations per joint angular bin, $N_k$, with consistent performance as $N_k$ varies from 5 to 20. However, many angular bins remain empty due to the inherent sparsity of training samples in 3D space. During the inference stage, if a test sample falls within an unpopulated bin, the system defaults to the anechoic chamber patterns. 
A primary limitation of the proposed method lies in isolating the platform's structural influence from external environmental factors, such as reflections from nearby buildings or the ground. While the framework does not explicitly decouple these components, it captures the aggregate wireless characteristics, which is beneficial as long as predictive performance improvement is concerned. 
To evaluate the impact of system bandwidth, Table~\ref{tab:per_compr2} summarizes the performance gains across various configurations. The average gain is consistent across 1.25~MHz, 2.5~MHz, and 5~MHz bandwidths within Dataset 2. 
MAE improvements for larger bandwidths are also evident for all scenarios. 
Finally, although this study focuses on LoS-dominant environments, the concept of platform-aware learning is extensible to non-line-of-sight (NLoS) scenarios. In such cases, the structural shadowing and multipath effects can be learned as a function of the nodes' intrinsic orientations (e.g., Euler angles).}
\section{Conclusion}
\EditMush{This letter introduced the concept of platform-aware 3D wireless link characterization, moving beyond traditional isolated antenna models to capture the complex structural scattering and reflection induced by node airframes. We formulated the \textit{mutual antenna pattern} as a joint function of the AoA and AoD, and demonstrated that this coupled pattern can be effectively learned from sparse, noisy field measurements using an LS framework. Our experimental validation using the AERPAW datasets shows that as few as $10$ measurements per joint angular bin are sufficient to achieve significant performance gains, reducing path loss estimation errors by up to $10$~dB compared to anechoic chamber baselines. These results underscore the importance of incorporating 3D terminal orientation into modern Channel Knowledge Maps (CKM). Future work will explore the extension of this mutual gain framework to multi-antenna systems and Non-Line-of-Sight (NLoS) environments by modeling link properties as a direct function of platform Euler angles.}
\bibliographystyle{IEEEtran}
\bibliography{ref}

\end{document}